\documentstyle [11pt,epsf] {article}

\begin{document}

\begin{center}
\vspace{2cm}
\LARGE
The Accretion of Gas onto Galaxies as Traced by their Satellites 
\\                                                     
\vspace{1cm} 
\large
Guinevere Kauffmann$^1$, Cheng Li$^1$, Timothy M. Heckman$^2$, \\                   
\vspace{0.5cm}
\small
\noindent
{\em $^1$Max-Planck Institut f\"{u}r Astrophysik, D-85748 Garching, Germany} \\
{\em $^2$Department of Physics and Astronomy, Johns Hopkins University, Baltimore, MD 21218}\\

\vspace{1.5cm}

\begin {abstract}
We have compiled a large sample of isolated central galaxies from the
Sloan Digital Sky Survey, which do not have a neighbour of 
comparable brightness within a projected distance of 1 Mpc. 
We use the colours, luminosities and surface brightnesses of 
satellite galaxies in the vicinity of these objects to estimate 
their neutral gas content and to derive the average total mass of HI gas contained in the satellites as a function of projected radius from the primary. Recent calibrations of merging timescales from N-body simulations are used to estimate the rate at which this gas will accrete onto the central galaxies. Our estimated accretion rates fall short of those needed to maintain the observed level of star formation in these systems by nearly two orders of magnitude. Nevertheless, there are strong correlations between the total mass of gas in satellites and the colours and specific star formation rates of central galaxies of all stellar masses. The correlations are much weaker if we consider the total stellar mass in satellites, rather than their total gas mass. We ask {\it why}  star formation in the central galaxies should be correlated with gas contained in satellites at projected separations of a Mpc or more, well outside the virial radius of the dark matter halos of these systems. We suggest that gas-rich satellites trace an underlying reservoir of ionized gas that is accreted continuously, and that provides the fuel for ongoing star formation in galaxies in the local Universe.

\end {abstract}
\end{center}
\vspace{1.5cm}
\normalsize
\pagebreak

\section {Introduction}

Our current theoretical understanding of how galaxies form through the
hierarchical formation of structure and through gas-dynamical dissipative
processes (White \& Rees 1978) leads to the view that star-forming
spiral galaxies like our own Milky Way must continue to accrete gas
at the present day. The observed size distribution of  disk galaxies
is consistent with models in which the specific angular momenta of the
disks are similar to those of their present-day halos and in which the
disks  assemble recently ($z<1$ on average) (Mo, Mao \& White 1998). The
fraction of baryons in dark matter halos that have condensed into galaxies
reaches a maximum of $\sim 20$ \% at masses close to that of the Milky Way
(Guo et al 2010). Unless the missing baryons have been ejected far from
the halo by some catastrophic feedback mechanism, gas is predicted to
accrete onto our Galaxy at the present epoch at average rates of around
a few solar masses per year (e.g. Kauffmann, White \& Guiderdoni 1993;
De Lucia \& Helmi 2008).

Observational support for continued accretion of gas in present-day disk
galaxies comes from studies of their stellar populations, which indicate
that star formation rates in the solar neighbourhood have remained
fairly constant over much of the age of the Universe (Binney, Dehnen \&
Bertelli 2000). In addition, chemical evolution models of the Galaxy
invoke continued gas infall and long formation timescales of the thin
disk to explain chemical abundance ratios in disk stars (e.g. Chiappini,
Matteucci \& Gratton 1997). However, {\em direct} observational evidence
for continued gas accretion remains elusive.  In a recent review article,
Sancisi et al (2008) provide a list of 24 nearby galaxies that appear
to be interacting with companions with substantial HI gas. These authors
estimate that around 25\%  of all spiral and irregular galaxies
show evidence for such minor interactions. A very rough estimate of the
resulting accretion rate is $\sim 0.1-0.2 M_{\odot}$ yr$^{-1}$, which
is too low to maintain star formation in the Milky Way at its observed
level of around 2-3  $M_{\odot}$ yr$^{-1}$.

A similar argument that the main gas reservoir that fuels ongoing star
formation in galaxies cannot be in the form of neutral hydrogen has
been made in a paper by Hopkins et al (2008).  These authors compare
the cosmic evolution of the star formation rates in galaxies with that
of their neutral hydrogen densities. Because the average star formation
rates in galaxies drop steeply from high redshift to the present day, but
the total neutral hydrogen density remains approximately constant with
redshift, they conclude that HI must be continuously replenished from
an external reservoir. This external reservoir may be in the form of a
hot galactic corona in hydrostatic equilibrium with the surrounding dark
matter halo.  Thermal instabilities result in the formation of clouds of
colder gas, which then rain onto the galactic disk  (e.g. Peek, Putman
\& Sommer-Larsen 2008).  Alternatively, the reservoir may reside in
filaments formed during the non-linear collapse of structure predicted
in CDM cosmologies (e.g. Keres et al 2005; Dekel et al 2009).  In this case, one might
expect  galaxies containing significant neutral hydrogen to {\em trace}
the underlying reservoir of ionized gas, because these galaxies would be
located in the higher density collapsed halos that make up the filaments.

A statistical analysis of how the properties of companion (or satellite)
galaxies relate to the properties of their hosts is only possible using
large optical imaging and spectroscopic surveys such as the Sloan Digital
Sky Survey (SDSS; York et al. 2000) or the Two Degree Field
 Survey (2dF; Colless et al 2001). Traditionally,
the average number density profiles and line-of-sight velocity dispersions
of satellites are used to place constraints on the density profiles of the
dark matter halos surrounding their  hosts (Zaritsky \& White 1994; Prada
et al 2003).  A few studies have examined whether there are correlations
between  satellites and hosts in terms of directly observable  quantities
such as colour or morphological type.  In particular, Weinmann et al
(2006) found that the  fraction of early-type satellites is significantly
higher in a halo with an early-type central galaxy than in a halo of the
same mass with a late-type central galaxy. They dubbed this phenomenon
``galactic conformity'', but did not come up with a compelling explanation
for why it should occur.

In this paper, we revisit the connection between satellite galaxies
and their hosts.  We use the colour, luminosity and surface brightness
of each satellite to estimate both its stellar mass and its neutral gas
fraction using empirical relations from the literature. We stack together
host galaxies with similar properties (such as stellar mass and colour)
and compute the total stellar and gas mass contained in satellites as a
function of projected radius $R_p$.  Finally, we study how these average
stellar and gas mass profiles depend on the properties of the host, and
we argue that the ``galactic conformity'' phenomenon can be understood
in terms of  ongoing fuelling of the interstellar medium of galaxies
by a reservoir of gas that is associated with their  satellites.

Our paper is organized as follows. In section 2, we describe the
spectroscopic and photometric galaxy samples used in our analysis, as
well as our methodology for calculating the average radial profiles of
the total  stellar and gas mass contained in satellites. In section 3,
we present the radial profiles as a function of the stellar mass of
the host and we confirm that the total amount of neutral gas contained
in the satellites falls  far short of what is needed to maintain
the observed levels of star formation in the hosts.  In section 4,
we examine correlations between the total amount of neutral gas and
stellar mass in the satellites and properties of the hosts, such as
colour,  concentration index and stellar surface
mass density. We demonstrate that the primary correlation is between
the total amount of gas in the satellites and the stellar populations
of the hosts. In section 5, we present what we believe to be the most
likely interpretation of our results. We assume a $\Lambda$CDM cosmology with
$\Omega_m$=0.3, $\Omega_{\Lambda}=0.7$ and $h=0.7$ throughout.

\section {The Data}

\subsection {The Photometric Sample of Satellites} Our sample is
constructed using the  ``datasweep'' files included as part of the  
New York University Value Added Catalogue (NYU-VAGC; Blanton et al. 2005).
This is a compressed version of the full photometric catalogue of the
SDSS Data Release 7 (Abazajian et al. 2009) that was used by Blanton et
al. to build the NYU-VAGC. The reader is referred to Blanton et al. (2005)
for a detailed description of the NYU-VAGC. More information about the
datasweep files is also available at  http://sdss.physics.nyu.edu/vagc/.
There are two catalogues in the datasweep, one for stars and one for
galaxies. Starting from the galaxy catalogue, we select all galaxies with
$r$-band apparent model magnitudes in the range $10<r<21$ after correction
for Galactic extinction, and PSF and model magnitudes satisfying $m_{psf}
- m_{model} > 0.145$ in all  five bands.  In order to select unique
objects in a run that are not at the edge of field, we require the 
{\tt RUN PRIMARY} flag  to be set, and the  {\tt RUN EDGE} flag not to be set.
This procedure
results in   a  final  sample  of $\sim  26$  million galaxies.

\subsection {The Spectroscopic Sample of Primaries} The parent sample
of primary galaxies for this study is composed of 397,344 objects
which have been spectroscopically confirmed as galaxies and have data
publicly available in the SDSS Data Release 4 (Adelman- McCarthy et al
2006). These galaxies are part of the SDSS main galaxy sample used for
large scale structure studies (Strauss et al 2002) and have Petrosian r
 magnitudes in the range $14.5 < r < 17.77$
after correction for foreground galactic extinction using the reddening
maps of Schlegel, Finkbeiner \& Davis (1998). Their redshift distribution
extends from  0.005 to 0.30, with a median redshift  of 0.10. Stellar
masses, metallicity and star formation rate estimates, as well as
spectral indices such as the 4000 \AA\ break strengths are available
at http://www.mpa-garching.mpg.de/SDSS/DR4/ (see Kauffmann et al 2003;
Brinchmann et al 2004; Tremonti et al 2004 for more details). In this
analysis we, further limit the spectroscopic sample  to galaxies with
$r < 17$, so that we are able to probe satellites down to a limiting
magnitude that is 4 magnitudes fainter than the primary (i.e. roughly
a factor 40 smaller in stellar mass).  We also restrict the sample to
lie at redshifts  $z>0.03$, so that we are able to probe satellites to reasonably
large projected radii from the primary object. Finally, we  eliminate galaxies
that lie at
projected radii of less than 1 Mpc from the photometric survey boundaries.

\subsection {Estimating the Stellar and Gas Masses of the Satellites}

We begin by finding all galaxies in the photometric sample that lie
within 1 Mpc in projected radius from the primary. The projected radius is
calculated by placing the satellite at the same redshift as the primary
($z_{prim}$). Some galaxies will  be interlopers that  are actually  at a very different
redshift, but we will  correct
the stellar and gas mass profiles for this effect via statistical background subtraction.

The next step is to estimate k-corrections and stellar $M/L$ ratios for
the satellites. Following the
procedure outlined in Bell et al (2003),  we compare the $gri$ fluxes of the galaxies
with a set of stellar population synthesis (SPS) models (we do not use
the $u$ or $z$ band fluxes, because they have  much larger errors).
Bruzual \& Charlot (2003) stellar population 
synthesis models are used to  construct a set  of models
in which the star formation histories vary exponentially with time
and we tabulate the observed $g-r$ and $r-i$ colours of each model at
a set of different redshifts (0.03 to 0.3, in $\Delta z$ bins of 0.01).
If the colours of the  satellites  lie outside the range of colours predicted
by the model at $z=z_{prim}$, then the satellite is discarded from the
analysis. In practice, this procedure eliminates a large fraction of
the reddest galaxies with $r-i$ colours that imply  that they must lie at high redshifts
(Collister et al. 2007), and significantly improves
the $S/N$ of our mass profiles. For the galaxies that remain, we use the
best-fit model to derive k-corrected colours at $z=0$. These colours are
then used to estimate the stellar mass-to-light-ratio using the formula
$\log (M/L)_i= -0.222+0.864(g-r)_{(z=0)} -0.15$ given in Table 7 of
Bell et al (2003) for a Kroupa IMF. As discussed in Bell et al (2003),
typical $M/L$ ratio uncertainties are expected to be $\sim 0.1$ dex.

In a recent paper, Zhang et al. (2009) used a sample of 800 galaxies
with HI mass measurements from the HyperLeda catalogue and optical
photometry from SDSS to calibrate a new photometric estimator of
the HI-to-stellar-mass ratio for nearby galaxies. The estimator is $
\log(GHI/S) = -1.73238(g-r)_{(z=0)} + 0.215182\mu_i - 4.08451$, where
$\mu_i$ is the i-band surface brightness estimated within the half-light
radius of the galaxy .  The estimator was shown to have a scatter of
0.31 dex in $\log (GHI/S)$. The authors did not find any significant
dependence of the residuals on  galaxy properties such
as stellar mass or mean stellar age (as measured by the 4000\AA\ break
strength). In this paper, we use this estimator to calculate an HI mass
for each of our satellite galaxies.

\subsection {Correction for interlopers}

We have generated 20 catalogs of galaxies that are spread over the
same area of sky as our sample of primary galaxies, but with randomly
generated sky coordinates. The galaxies in the random catalogues are
assigned the redshifts and IDs of the galaxies in our primary sample,
and we proceed to find ``satellites'' in the photometric catalogue in
exactly the same way as for the real sample of primaries. We also apply
exactly the same procedures to derive  k-corrections, to throw out
galaxies that do not fit in the predicted colour range, and to derive
stellar and gas masses. This allows us to estimate what fraction of the
stellar mass and gas mass at radius $R_p$ originates from interlopers,
rather than from true, physically-associated satellite systems.

Because the  contamination by interlopers at a given value of $R_p$ in
physical units will depend on the redshift of the primary galaxy, it is
important to compute the back ground using random galaxies with exactly
the same distribution of redshifts as that of the real primary galaxies
in our stacks. Our procedure, therefore, can be summarized as follows:
\begin {enumerate} \item  Partition the sample of primary  galaxies into
bins of  a given
   property (e.g. stellar mass).
\item Find  satellites within 1 Mpc of each primary and  estimate
  stellar and gas masses.
\item  Compute the average stellar and HI mass
  profiles as a function of projected radius for the sum of all primaries
  in   given  bin.
\item Repeat steps 2 and 3 above for the corresponding set of
     randomly distributed primaries.
\item The final stellar mass or HI mass profile for galaxies in a
given bin is found by subtracting the average profiles obtained for the
randomly distributed primaries from the profiles derived in step 3 above.
\end {enumerate}

\section {Results}

\subsection {Stellar and Gas Mass Profiles}

In Figure 1, we show plots of the cumulative stellar and gas mass
contained in satellites interior to a projected radius $R_p$. Results
are shown in bins of stellar mass. The red curves show the integrated
stellar mass profiles, while the blue curves show the integrated HI
gas mass profiles. To guide the eye, we re-plot the curves in the first
panel using dotted blue and red line styles.

\begin{figure}
\centerline{ 
\epsfxsize=9cm \epsfbox{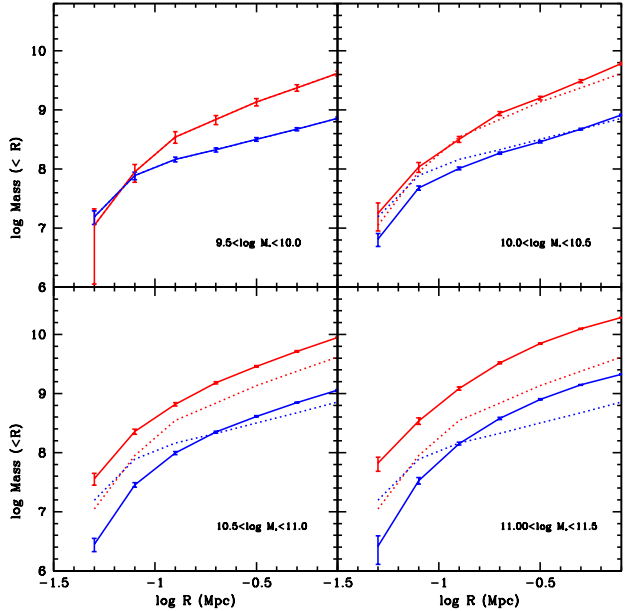}
}
\caption{\label{fig1}
\small
Cumulative stellar (red) and gas mass (blue) contained in satellites
interior to projected radius $R_p$. Results are shown in four different
bins of stellar mass. The dotted lines are plotted to guide the
eye and simply repeat the profiles in the first panel. }
\end {figure}
\normalsize

As can be seen, the total  stellar mass contained in satellites increases
as a function of the stellar mass of the primary. This increase is seen
at all value of $R_p$. However, the total mass  of gas contained
in satellites exhibits quite different behaviour as a function of the
stellar mass of the primary galaxy.  On small scales, the mass in gas actually
{\em decreases} as a function of increasing host galaxy mass.

One problem with simply stacking together all galaxies in bins of stellar
mass, is that one is effectively averaging together galaxies in many
different environments.  In particular, a galaxy with a stellar mass of
a few $\times 10^{10} M_{\odot}$ could  be a central galaxy in a halo
of mass $\sim 10^{12} M_{\odot}$, or it could be a satellite galaxy
in a much larger halo, with mass corresponding to that of a group or
cluster. Satellite galaxies in groups or clusters are known to have redder
colours and lower star formation rates than their counterparts in low
density environments. It is hypothesized that the main reason for this is
that the external gas reservoirs of satellites are stripped as they orbit
within the cluster potential. Because the main goal of this study is to
understand if there is a link between star formation in a galaxy and the
gas contained in satellites, it is important to remove primary galaxies
that are themselves satellite systems in massive dark matter halos.

We do this by imposing an isolation criterion on our sample of primaries.
We exclude all galaxies with a neighbour with an apparent magnitude
brighter than $r_{prim}+0.4$ mag (where $r_{prim}$ is the r-band model
magnitude of the primary) within a projected radius of 1 Mpc from
the primary. This is a fairly stringent criterion that should
eliminate all galaxies located in groups and clusters. \footnote {Note that
we have experimented with different isolation criteria, and none of our conclusions
would change if we adopted a different cut.}  The stellar mass and gas mass profiles of this sample of
isolated primaries is shown in Figure 2. We overplot the mass profiles
for the full samples as dashed lines in order to show the effect of
the isolation criterion. As can be seen, the cut has the largest effect
on the low mass primaries. For galaxies with stellar masses less than
a few $\times 10^{10} M_{\odot}$, the stellar mass contained within a
projected radius of 1 Mpc drops by an order of magnitude, showing that
most of the contribution is coming from massive halos.  On the other
hand, the gas mass within 1 Mpc drops by less than a factor of two,
presumably because most of the galaxies in massive halos are red systems
with small HI fractions.  For galaxies with stellar masses greater than
$10^{11} M_{\odot}$, the isolation criterion has very little effect,
because most of these galaxies are the central galaxies of their halo.
We have also marked the expected location of the virial radius of the dark matter
halos of the galaxies in each stellar mass bin. We adopt the standard 
definition of the  virial radius 
as the radius within which the dark matter over-density is 200 times the critical density,
and we use the empirically-calibrated  relation between the stellar mass of a central
galaxy  and its  dark matter halo mass
given in Table 2 of Wang et al (2006).

\begin{figure}
\centerline{ 
\epsfxsize=9cm \epsfbox{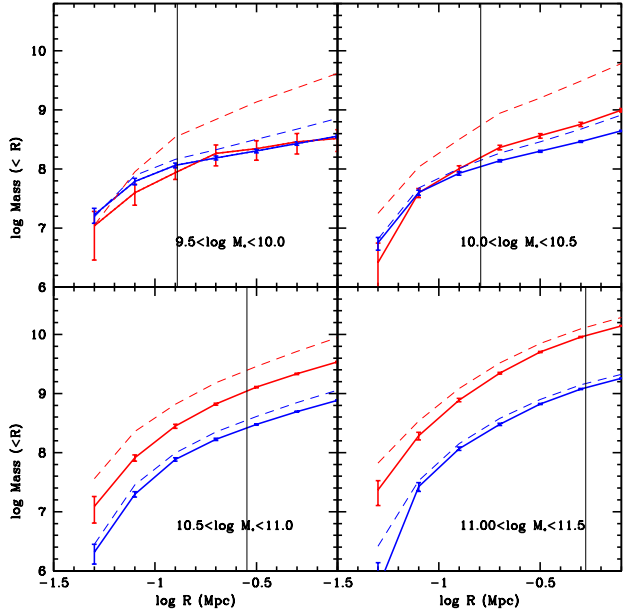}
}
\caption{\label{fig2}
\small
Cumulative stellar (red) and gas mass (blue) contained in satellites
interior to projected radius $R_p$ around ``central'' galaxies (see text).
The dashed lines show the results obtained for all galaxies, 
as plotted in Figure 1. The vertical black lines show the 
expected location of the virial radius of a dark matter halo
hosting a galaxy in the given stellar mass range (see text).}
\end {figure}
\normalsize

\subsection {The inferred gas accretion rate}

We now estimate the average rate at which gas contained in satellites will
accrete onto central galaxies as a function of their mass, and compare
the estimated gas accretion rate with the average star formation rates
in the primaries. To do this, we make use of the calibration between the
projected separation of a pair of galaxies and the average time until the
merger given in  Kitzbichler \& White (2008).  These authors use virtual
galaxy catalogues derived from the Millennium Simulation (Springel et al
2005) to derive sample-averaged merging times as a function of projected
separation, the stellar masses of the galaxies in the pair, and  redshift. At
$z<2$ , the timescale  $T$ is only weakly dependent on mass and can be
approximated as $T \sim T_0 r_{25} M_{*}^{-0.3}$, where $r_{25}$ is the
projected separation in units of 25 $h^{-1}$ kpc and the coefficient $T_0$
is 1.6 Gyr (appropriate for samples where the line-of-sight velocity
difference between the two galaxies is unconstrained).

In the left panel of  Figure 3, we plot the average star formation rates
of the central galaxies in our sample for 5 different bins in stellar
mass.  The star formation rates are estimated using  emission line fluxes
and ratios measured from the SDSS spectra  and corrected for aperture
effects as described in Brinchmann et al (2004).  The star formation
rates rise from $\sim 1.5 M_{\odot}$ yr$^{-1}$ for central galaxies of
$10^9 M_{\odot}$ to $\sim 3 M_{\odot}$ yr$^{-1}$ for central galaxies
like the Milky Way with stellar masses of $3 \times 10^{10}-10^{11}
M_{\odot}$. In the middle panel, we plot the merging time as a function of
projected radius $R_p$ given by the calibration of Kitzbichler \& White
(2008). Blue, green, black, red and magenta lines show results for the
five mass bins plotted in the left panel, with stellar mass increasing
as colour goes from blue to magenta. Finally, in the right panel we plot
the fraction of the total star formation rate of the primary galaxy that
is contributed from gas  accretion in satellites interior to
radius $R_p$, i.e. $M_{gas}(r<R_p)/T(R_p)/SFR_{primary}$.

\begin{figure}
\centerline{ 
\epsfxsize=14cm \epsfbox{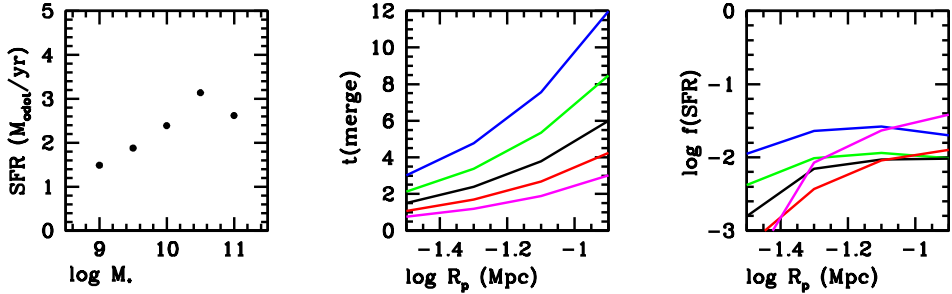}
}
\caption{\label{fig3}
{\em Left}: Average star formation rates
of the central galaxies in our sample for 5 different bins in stellar
mass. {\em Middle:} Merging time as a function of
projected radius $R_p$ as given by the calibration of Kitzbichler \& White
(2008). Blue, green, black, red and magenta lines show results for the
five mass bins plotted in the left panel, with stellar mass increasing
from blue to magenta. {\em Right:} 
The fraction of the total star formation rate of the primary galaxy that
is contributed from accreted  gas in satellites interior to
radius $R_p$.}
\small
\end {figure}
\normalsize

It is interesting that for low mass primaries, the curves are reasonably
flat, so it does not matter what radius one adopts to estimate the gas
accretion rate. For high mass primaries, the estimated accretion rate
drops at small projected separations. It is reasonable  that not all the
gas present in a satellite  far  from the primary will remain available
to fuel star formation by the time that satellite has finally reached
the centre of the halo. Part of the gas may be consumed into stars or
stripped out of the galaxy. However, we caution against drawing too many
strong conclusions about satellite stripping processes from these result.
The merging times calibrated by Kitzbichler \& White are not guaranteed
to remain accurate at very small pair separations, because the resolution
of the Millennium Simulation was not high enough to track the orbital
evolution  of merging galaxies in detail. The main conclusion from Figure
3 is that the estimated gas accretion rates from satellites always fall
short of what is needed to maintain the observed level of star formation
in the primary galaxies by around two orders of magnitude. This shortfall
is roughly independent of the mass of the primary.

\section {Correlations with  properties of the primaries}

In this section we examine whether the integrated stellar mass and gas
mass profiles of satellites correlate with the properties of the primary.
In Figure 4, we split the our primary sample into ``blue sequence''
galaxies with $g-r<0.65$ and ``red sequence'' galaxies with $g-r>0.65$
\footnote {Note that the division between the red and blue sequence
moves to slightly redder colours at higher stellar masses, but for
simplicity we choose a fixed cut in colour}  and we plot the stellar
and gas mass profiles for four different bins in the stellar mass of
the primary. Interestingly, the stellar mass profiles of the satellites
do not correlate with the colours of the central galaxies, but there is
significantly more gas present in the satellites around the blue central
galaxies. The effect is strongest st small radii, but persists
out to projected distances of $\sim 1$, except in our highest
stellar mass bin.   Note that the fraction of blue central galaxies is a very strong
function of stellar mass. In the lowest mass bin ($9.5<\log M_*<10$),
there are too few red galaxies to include in the plot,  and in our
highest mass bin ($11.0<\log M_*<11.5$), blue galaxies comprise only a few percent
of the sample. Nevertheless, the strong systematic difference in the gas
profiles around red and blue galaxies is present at all stellar masses .

\begin{figure}
\centerline{ 
\epsfxsize=12cm \epsfbox{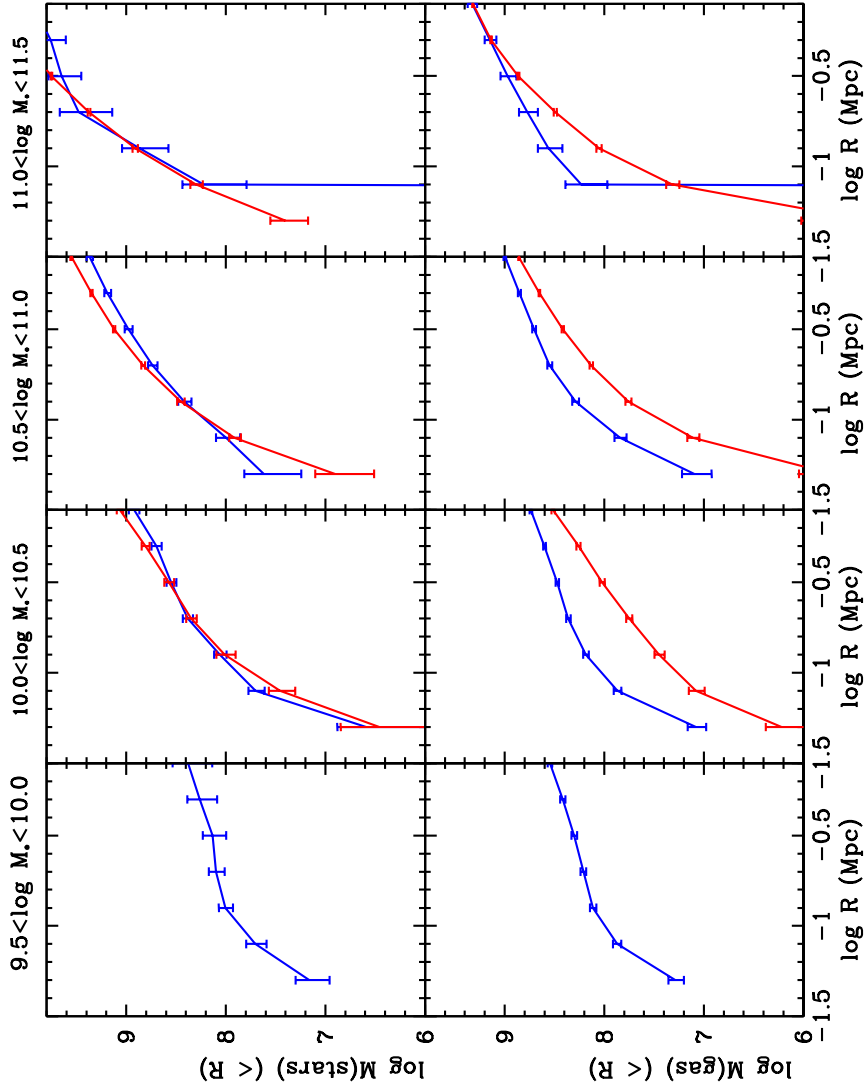}
}
\caption{\label{fig2}
\small
Cumulative stellar (top) and gas mass (bottom) contained in satellites
interior to projected radius $R_p$ around ``central'' galaxies.
The central galaxy population is divided into blue sequence galaxies
with $g-r<0.65$ (blue curves) and red sequence 
galaxies with $g-r>0.65$ (red curves).}
\end {figure}
\normalsize

Figure 5 expands upon these results by showing what happens if we further
divide galaxies by colour. The thick red and blue curves  repeat the
results shown in Figure 4. The green and cyan curves are for galaxies
with $g-r<0.45$ and $g-r<0.35$, respectively. The magenta curves are
for galaxies with $g-r > 0.75$.  As can be seen, the colours of red
sequence galaxies do not appear to be related to either  the stellar
mass or the gas mass profiles of the satellites in their vicinity.
However, on the blue sequence there is a clear and monotonic trend for
the amount of gas to increase as the colour of the primary becomes bluer.
For the bluest primaries, one also sees an increase in the stellar mass
contained in satellites, but the effect is weaker than that seen
for the gas.  An increase in the {\em number} of satellites around the
most strongly star-forming galaxies is to be expected. Previous work
has shown that that most strongly star-forming galaxies exhibit an
excess of close pairs (Barton, Geller \& Kenyon 2000; Li et al 2008),
and this is generally understood as evidence that tidal interactions
between galaxies induce higher rates of star formation in these objects.
What is surprising about our results, is that if one re-expresses the
traditional statistics of close pairs in terms of the {\em gas content}
of the nearby satellites, the trends become much stronger and also extend
over the whole blue sequence, instead of just being apparent for the
most strongly star-bursting galaxies.  We will come back to this point
in the discussion.

Finally, we note that the correlation between the colour of the primary and the
gas contained in satellites is not a phenomenon confined to satellites at small   
spatial separations from the primary ; Figure 5 shows the 
correlation persists beyond  scales of $\sim 1$ Mpc, i.e. 
beyond the virial radius of the dark matter halo 
for  the majority of the primaries in our sample. 
Only for the the most massive primaries with stellar masses
$\sim 10^{11} M_{\odot}$, is there some evidence that the correlation is confined
to satellites within the virial radius of the dark matter  halo.

\begin{figure}
\centerline{ 
\epsfxsize=12cm \epsfbox{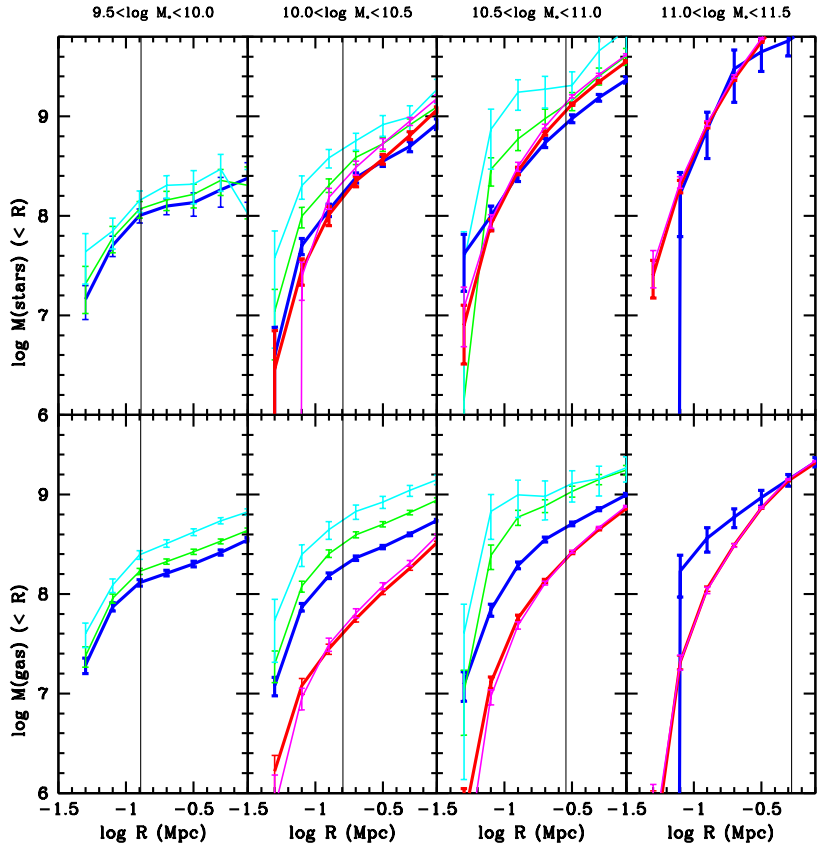}
}
\caption{\label{fig2}
\small
Cumulative stellar (top) and gas mass (bottom) contained in satellites
interior to projected radius $R_p$ around ``central'' galaxies.
The central galaxy population is divided into blue sequence galaxies
with $g-r<0.65$ (blue curves) and red sequence 
galaxies with $g-r>0.65$ (red curves) In addition, the blue sequence
in further subdivided into  galaxies with $g-r<0.45$ (green)
and $g-r<0.35$ (cyan). Magenta curves show the very reddest galaxies
with $g-r > 0.75$.
The vertical black lines show the 
expected location of the virial radius of the dark matter halos that host
galaxies in the given stellar mass range.}
\end {figure}
\normalsize

\subsection {Star formation or morphology?}

The results discussed above have been presented in the literature before,
albeit in a different form.  Weinmann et al (2006) studied correlations
between the properties of satellites and central galaxies using data
from the Sloan Digital Sky Survey Data Release 2.  They found that the
"early-type" fraction of satellite galaxies is significantly higher in
a halo with an early-type central galaxy, than in a halo of the same mass
with a late-type central galaxy. They dubbed this phenomenon `galactic
conformity' and  considered  a variety of processes that might give rise to
the effect, including mergers, harrassment in clusters and ram-pressure
stripping, but concluded that none of these could provide a compelling
explanation for  the observational trends.

We note that the physical processes discussed in the Weinmann et al paper
are ones that would alter the {\em structural} properties of the galaxies,
but in actual fact  colours and spectral indices such
as the 4000 \AA\ break strength were used  to classify the galaxies in
the sample  by type.  It is well known that the star formation rates in
galaxies correlate strongly with their structural properties. Galaxies
with low stellar masses, stellar surface densities and concentrations tend
to be blue, while galaxies with high masses, stellar surface densities
and concentrations, tend to be red. This is why spectral type and
galaxy structure are frequently lumped together under the general term
"morphology". If one  wishes to understand whether a given  physical
process (e.g. gas accretion) could be  responsible for the correlation
between the properties of central galaxies and their satellites, it is
very important to decouple the star formation/galaxy structure link and
understand which of the two properties is {\em primarily} affected by
the phenomenon under consideration.

Our attempt to do this is shown in Figure 6. We plot contours of the
total gas mass in satellites estimated within a fixed  aperture of
$R_p=300$ kpc.  We show our results in the two-dimensional plane of $g-r$ colour versus
stellar surface mass density $\mu_*$ and concentration index (Concentration
index $C$ is defined
as the ratio between the radius enclosing  90\% of the total Petrosian
flux of the galaxy in the $r$-band and the radius enclosing 50\% of
this flux \footnote {As shown in Figure 1 of Weinmann et al (2009), the
concentration index correlates very well with the bulge-to-disk ratios
obtained by 2-dimensional multi-component fits of Gadotti (2008) using
the BUDDA code.}) Results are shown for two different bins in stellar mass.
As can be seen,  there is a strong correlation between
central galaxy colour and both surface density and concentration,
so that galaxies in our sample only occupy a relatively narrow  strip
in the plots. We can also see that the contours of constant gas mass
in satellites always run parallel to the  x-axis in all four panels,
showing that the main link is between gas mass in satellites and  the colour of the
central galaxy,  and not between gas mass and the structure of the central galaxy.

\begin{figure}
\centerline{ 
\epsfxsize=12cm \epsfbox{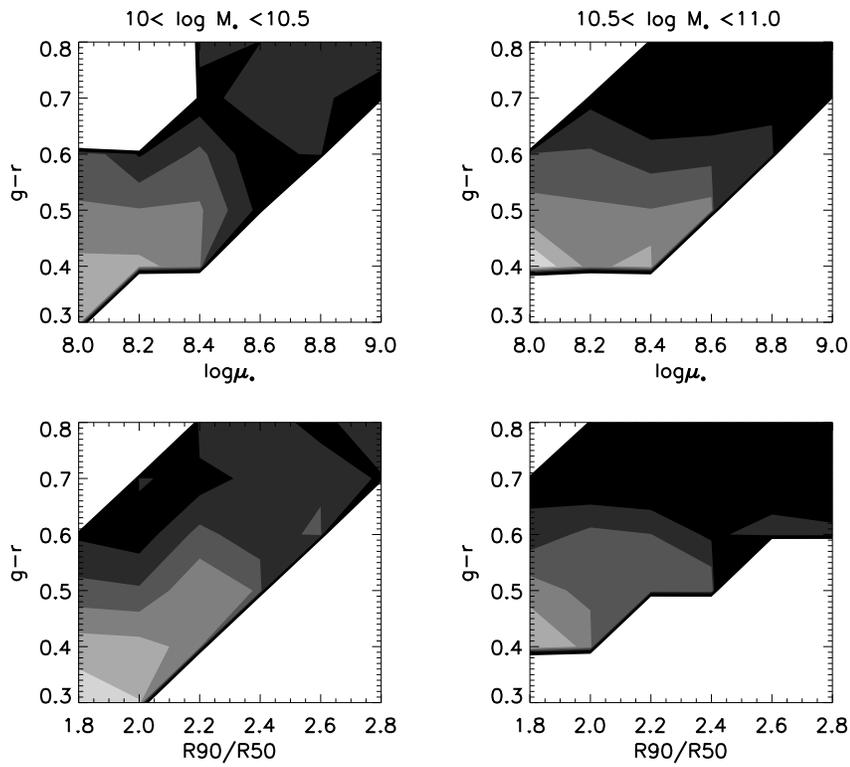}
}
\caption{\label{fig6}
\small
Contours of the
total gas mass in satellites estimated within a fixed  aperture of
$R_p=300$ kpc.  We show the results in the plane of $g-r$ colour versus
stellar surface mass density $\mu_*$ and concentration index, defined
as the ratio between the radius enclosing  90\% of the total Petrosian
flux of the galaxy in the $r$-band and the radius enclosing 50\% of
this flux.}
\end {figure}
\normalsize

\section {Discussion}

We now discuss the main conclusions and implications of this work.

{\bf 1. The cold gas contained in satellites is not sufficient to provide
the fuel for ongoing star formation in central galaxies.} As discussed
in the introduction, this conclusion is not a surprise. 
We note that our estimate of the accretion rate from gas
in satellites is  a factor of 2-5  lower than that of Sancisi et
al (2008). There are two likely reasons for this: a) we use merging
timescales from Kitzbichler \& White (2008), which are a factor of
2 longer than most other published determinations of merger
rates, b) the Sancisi et al study focuses on galaxies with significant HI
content, whereas our study averages over all galaxies of a given stellar
mass. As we have demonstrated, there is a strong {\em correlation} between
the colours (and hence gas content) of central galaxies and the mass of
gas in their satellites in the sense that bluer and more gas-rich galaxies
should have more gas in satellites in their immediate surroundings.

{\bf 2. Star formation in central galaxies is strongly correlated with the
mass of gas  contained in the satellites} It is not easy to understand
{\em why} the  properties of central and satellite galaxies that are
separated by distances of hundreds of kiloparsecs to several  Mpc ought
to correlate. One explanation  that should  be considered is that it is
the central galaxy that somehow affects the properties of is satellites.
Weinmann et al (2006) and Ann, Park \& Choi (2008) argue  that the satellites
of early-type central galaxies are likely to be deprived of their  gas
reservoirs through  hydrodynamic interactions with the X-ray
emitting hot gas of their host. 

 We do not think this is a viable
explanation of our results, because we have shown that the correlation
between star formation in the host and gas in the satellites extends
down to central galaxies with stellar masses of a few  $ \times 10^{9}
M_{\odot}$. These galaxies are not massive enough to have hot gas
halos. In addition, the correlations are only present for galaxies on
the blue sequence; there is no apparent relation between the colours of
galaxies on the red sequence and the gas contained in 
their satellites.

The hypothesis that we deem  more likely is that the satellites {\em
trace a much larger underlying reservoir of  ionized gas} that is
accreting onto the central galaxies in a roughly continuous fashion. N-body plus
hydrodynamical simulations of structure formation assuming a standard
$\Lambda$CDM cosmology show that baryons in  the $z=0$
universe are partitioned relatively evenly between a condensed phase
consisting of stars and cold galactic gas, a diffuse phase consisting of
photo-ionized intergalactic gas with temperatures below $10^5$ K,
and the so-called warm-hot intergalactic medium (WHIM) consisting
of  collisionally ionized gas  at
temperatures of between $10^5$ and $10^7$ K (e.g. Cen \& Ostriker 1999;
Dave et al 2001; Kang et al 2005). Most of  the WHIM gas
is  too diffuse to self-gravitate or self-shield, 
and thus cannot achieve the densities required
to make radiative cooling processes important (Dave et al 2001).
Detailed studies of gas accretion onto galaxies with 
hydrodynamical simulations show that much of the gas reservoir for star
formation is in the form of $10^4-10^6 K$ gas associated with filaments
with coherence length of greater than 1 Mpc (Keres et al 2009).

These studies  then suggest that the satellite galaxies analyzed in this paper 
trace the high density peaks of  an underlying
cosmic web of mostly photo-ionized gas at reasonably high overdensities,
and that this ionized gas is the
primary reservoir for future star formation in the galaxy. 
 If there is  more gas in  satellites in the
vicinity of a given galaxy, there is also likely to be more diffuse gas between the
satellites. This would explain why there are such strong correlations
between star formation and 
cold gas in satellites that are at distances of a Megaparsec or more from
the primary, where infall times are long and the gas in the satellites themselves could 
never have has any physical contact with the primary galaxy. 

{\bf 3. The same correlation is seen for galaxies with stellar masses greater
than $10^{11} M_{\odot}$.} Mandelbaum et al (2006) have determined the
average relation between the stellar mass of a galaxy and the mass of its
dark matter halo from an analysis of galaxy-galaxy weak lensing in the
SDSS. Late-type galaxies with masses of $10^{11} M_{\odot}$ were found to
reside in dark matter halos of mass $ \sim 2 \times 10^{12}  M_{\odot}$,
while early-type galaxies of the same stellar mass were located in
somewhat more massive halos ($ \sim 5 \times 10^{12} M_{\odot}$).
Hydodynamical simulations show that atmospheres of hot virialized gas
develop in halos more massive than $2-3 \times 10^{11} M_{\odot}$ and
this  transition mass remains nearly constant with redshift (e.g. Keres
et al 2005, 2009). 

One important question is whether star formation in the most massive
galaxies  is fuelled by gas that accretes from the hot phase. Models have
been proposed (Maller \& Bullock 2004; Kaufmann et al 2006;  Peek, Putman
\& Sommer-Larsen 2008)  in which ongoing accretion onto galactic disks
progresses via the unstable cooling of the baryonic halo into condensed
clouds, which then rain onto the disk. If this is the main mode by which
gas reaches central galaxies in massive halos, one would no longer expect
to see any correlation between star formation in these objects and the
amount of gas contained in satellites.  The fact that the correlation
persists for the most massive galaxies in our sample supports
the hypothesis that blue massive galaxies are found in a {\em small subset}
of halos where the  gas reservoir  is still spatially linked to
satellite galaxies with neutral gas. The reason why blue
massive galaxies are rare is because the majority of massive halos have
completely transitioned to supporting  hot gas atmospheres, which are
unable to cool efficiently and form stars

Although we believe that the picture put forward in this concluding
discussion is a plausible one, detection of the underlying  reservoir  will be required
before we can be sure that we have located the  fuel supply for
ongoing star formation in nearby galaxies. Unfortunately the temperature
and low density of diffuse gas around galaxies  leave it nearly undetectable in emission.
The most direct evidence we have that this component exists comes
from studies of higher ionization lines of oxygen  detected in
absorption against background quasars (e.g. Richter et al 2004). These
observations only probe individual sightlines and cannot accurately map
the distribution and extent of the cosmic web. Nevertheless,  if one has
available a large enough sample of sightlines that pass within a few
Mpc of  nearby galaxies,  one might nevertheless expect to see strong
statistical correlations between star formation in the galaxy and the
inferred column density of warm hydrogen, similar to the correlation
we have found  between star formation and the  total gas contained in
satellites. This may be possible in future with the Cosmic Origins
Spectrograph (COS) on board the Hubble Space Telescope. 
Eventually, imaging of the IGM through its emission in
Ly$\alpha$  may become possible with a wide-field 
UV spectral imaging telescope in space
(Sembach et al 2009), or  with next generation X-ray observatories that
will be able to detect higher temperature gas (Bregman et al 2009).

\vspace{1.2cm}
\large
\noindent
{\bf Acknowledgements}\\
\normalsize

\noindent

Funding for the creation and distribution of the SDSS Archive has been 
provided by the Alfred P. Sloan Foundation, the Participating Institutions, 
the National Aeronautics and Space Administration, 
the National Science Foundation, the U.S. Department of Energy, 
the Japanese Monbukagakusho, and the Max Planck Society. 
The SDSS Web site is http://www.sdss.org/.
The SDSS is managed by the Astrophysical Research Consortium (ARC) 
for the Participating Institutions. The Participating Institutions 
are The University of Chicago, Fermilab, the Institute for Advanced Study, 
the Japan Participation Group, The Johns Hopkins University, 
the Korean Scientist Group, Los Alamos National Laboratory, 
the Max-Planck-Institute for Astronomy (MPIA), 
the Max-Planck-Institute for Astrophysics (MPA), 
New Mexico State University, University of Pittsburgh, 
University of Portsmouth, Princeton University, 
the United States Naval Observatory, and the University of Washington.

\pagebreak
\Large
\begin {center} {\bf References} \\
\end {center}
\normalsize
\parindent -7mm
\parskip 3mm

Abazajian K.~N., et al., 2009, ApJS, 182, 543

Adelman-McCarthy J.~K., et al., 2006, ApJS, 162, 38

Ann H.~B., Park C., Choi Y.-Y., 2008, MNRAS, 389, 86 

Barton E.~J., Geller M.~J., Kenyon S.~J., 2000, ApJ, 530, 660 

Bell E.~F., McIntosh D.~H., Katz N., Weinberg M.~D., 2003, ApJS, 149, 289 

Blanton M.~R., et al., 2005, AJ, 129, 2562 

Binney J., Dehnen W., Bertelli G., 2000, MNRAS, 318, 658 

Bregman J.~N., et al., 2009, astro, 2010, 24

Brinchmann J., Charlot S., White S.~D.~M., 
Tremonti C., Kauffmann G., Heckman T., Brinkmann J., 2004, MNRAS, 351, 1151 

Bruzual G., Charlot S., 2003, MNRAS, 344, 1000 

Cen R., Ostriker J.~P., 1999, ApJ, 514, 1 

Chiappini C., Matteucci F., Gratton R., 1997, ApJ, 477, 765 

Colless M., et al., 2001, MNRAS, 328, 1039 

Collister A., et al., 2007, MNRAS, 375, 68 

Dav{\'e} R., et al., 2001, ApJ, 552, 473 

De Lucia G., Helmi A., 2008, MNRAS, 391, 14 

Dekel A., et al., 2009, Natur, 457, 451 

Gadotti D.~A., 2008, MNRAS, 384, 420 

Guo Q., White S., Li C., Boylan-Kolchin M., 2010, MNRAS, 367 

Hopkins A.~M., McClure-Griffiths N.~M., Gaensler B.~M., 2008, ApJ, 682, L13 

Kang H., Ryu D., Cen R., Song D., 2005, ApJ, 620, 21 

Kauffmann G., White S.~D.~M., Guiderdoni B., 1993, MNRAS, 264, 201 

Kauffmann G., et al., 2003, MNRAS, 346, 1055

Kaufmann T., Mayer L., Wadsley J., Stadel 
J., Moore B., 2006, MNRAS, 370, 1612 

Kere{\v s} D., Katz N., Weinberg D.~H., 
Dav{\'e} R., 2005, MNRAS, 363, 2 

Kere{\v s} D., Katz N., Dav{\'e} R., 
Fardal M., Weinberg D.~H., 2009, MNRAS, 396, 2332 

Kitzbichler M.~G., White S.~D.~M., 2007, MNRAS, 376, 2 

Li, C., Kauffmann G., Heckman T.~M., Jing Y.~P., White S.~D.~M., 2008, MNRAS, 
385, 1903 

Maller A.~H., Bullock J.~S., 2004, MNRAS, 355, 694 

Mandelbaum R., Seljak U., Kauffmann G., 
Hirata C.~M., Brinkmann J., 2006, MNRAS, 368, 715 

Mo H.~J., Mao S., White S.~D.~M., 1998, MNRAS, 295, 319 

Peek J.~E.~G., Putman M.~E., Sommer-Larsen J., 2008, ApJ, 674, 227 

Prada F., et al., 2003, ApJ, 598, 260 

Richter P., Savage B.~D., Tripp T.~M., Sembach K.~R., 2004, ApJS, 153, 165

Sancisi R., Fraternali F., Oosterloo T., van der Hulst T., 2008, A\&ARv, 15, 189 

Schlegel D.~J., Finkbeiner D.~P., Davis M., 1998, ApJ, 500, 525 

Sembach K., et al., 2009, astro, 2010, 54 

Springel V., et al., 2005, Natur, 435, 629 

Strauss M.~A., et al., 2002, AJ, 124, 1810 

Tremonti C.~A., et al., 2004, ApJ, 613, 898 

Wang L., Li C., Kauffmann G., De Lucia G., 2006, MNRAS, 371, 537 

Weinmann S.~M., van den Bosch F.~C., Yang 
X., Mo H.~J., 2006, MNRAS, 366, 2 

White S.~D.~M., Rees M.~J., 1978, MNRAS, 183, 341 

York D.~G., et al., 2000, AJ, 120, 1579 

Zaritsky D., White S.~D.~M., 1994, ApJ, 435, 599 

Zhang W., Li C., Kauffmann G., Zou H., Catinella B., Shen S., Guo Q., Chang 
R., 2009, MNRAS, 397, 1243 

\end{document}